\documentclass[12pt]{article}

\usepackage{amsmath}
\usepackage{graphicx}
\usepackage{chngcntr}
\usepackage{textcomp}
\usepackage{rotating}
\usepackage{color}
\usepackage[table]{xcolor}
\usepackage{cite}

\usepackage{times}
\usepackage[markup=underlined]{changes}
\usepackage[english]{babel}

\graphicspath{{./figures/}}

\newenvironment{sciabstract}{%
\begin{quote} \bf}
{\end{quote}}

\title{{\bf Sculpting vesicles with active particles: Less is more}}

\author
{Hanumantha Rao Vutukuri$^{1,+,\ast}$, Masoud Hoore$^{2,3,+}$, Clara Abaurrea-Velasco$^{2,+}$, \\ 
Lennard van Buren$^{1}$, Alessandro Dutto$^{1}$, Thorsten Auth$^{2}$, \\
Dmitry A. Fedosov$^{2}$, Gerhard Gompper$^{2, \ast}$, Jan Vermant$^{1}$\\
\\
\normalsize{$^{1}$Soft Materials, Department of Materials, ETH Z\"urich, 8093 Z\"urich,  Switzerland}\\
\normalsize{$^{2}$Theoretical Soft Matter and Biophysics,}\\
\normalsize{Institute of Complex Systems and Institute for Advanced Simulation,}\\
\normalsize{Forschungszentrum J\"ulich, 52425 J\"ulich, Germany}\\
\normalsize{$^{3}$Department of Systems Immunology and Braunschweig Integrated Centre of Systems Biology,}\\
\normalsize{Helmholtz Centre for Infection Research, Braunschweig, Germany}\\
\\
\normalsize{$^+$ These authors equally contributed to this work;}\\
\normalsize{$^\ast$To whom correspondence should be addressed;}\\
\normalsize{E-mail: h.r.vutukuri@mat.ethz.ch,}\\ 
\normalsize{g.gompper@fz-juelich.de}
}

\date{}

\begin{document} 


\baselineskip24pt


\maketitle


\begin{sciabstract}
\newpage
\section*{Abstract} 
Biological cells are able to generate intricate structures and respond to external stimuli, 
sculpting their membrane from within  
\cite{Mattila_FMA_2008,Fletcher_CMC_2010,needleman2017active,turlier2016}. 
Simplified biomimetic systems can aid in understanding the principles which govern these 
shape changes and elucidate the response of the cell membrane under strong deformations 
\cite{Bausch-science,Dogtram-cell,Mulla_2018,petra-review2019,mellouli2013self}. 
Here, a combined experimental and simulation approach is used to identify the conditions 
under which different non-equilibrium shapes and distinct active shape fluctuations can be obtained by enclosing self-propelled 
particles in giant vesicles. Interestingly, the most pronounced shape changes are observed 
at relatively low particle loadings, starting with the formation of tether-like protrusions 
to highly branched, dendritic structures. At high volume fractions,  
globally deformed vesicle shapes are observed. The obtained state diagram of vesicles sculpted by active particles 
predicts the conditions under which local internal forces can generate dramatic 
cell shape changes, such as branched structures in neurons.
\end{sciabstract}


\section*{Introduction}
One of the fascinating aspects of biological cells is their ability to actively sense and 
respond to external stimuli, such as the extension of protrusions to explore their environment, 
and to experience non-equilibrium fluctuations \cite{Fletcher_CMC_2010,needleman2017active,turlier2016}. 
Such protrusions are generated by forces exerted on the cell's plasma membrane by the cytoskeleton \cite{Mattila_FMA_2008}, 
and even highly branched structures such as glial cells are encountered \cite{Andriezen_NGE_1893,Vasile_HAC_2017}. 
Listeria bacteria provide another fascinating example of showing how highly localized forces strongly deform the cell membrane that eventually leads to cell breakage when they infect mammalian cells \cite{theriot1992rate}. Minimal artificial soft matter model systems \cite{Bausch-science,Dogtram-cell,Mulla_2018,petra-review2019,mellouli2013self} that mimic certain features of biological systems may help us to understand  complex cellular phenomena \cite{petra-review2019,Mulla_2018}. One of the prominent questions in synthetic biology 
and soft robotics is whether and how such soft artificial systems 
with highly localized internal forces from within, mimicking some aspects of biological cells, can be engineered. One key aspect is to understand how to generate and control shape changes from inside. In this work, we present a simplified biomimetic experimental model system in order to address the question how cells reconfigure their shape and how the membrane responds to a localized point forces from within, such as those exerted by the cytoskeleton. In our combined experimental and simulation study, the plasma membrane of biological cells 
is mimicked by the lipid bilayer of giant unilamellar vesicles, and the local internal forces are generated by enclosed 
self-propelled colloidal particles. 

In recent years, self-propelled particles (SPPs) have emerged as versatile model systems to study generic aspects of active, out-of-equilibrium systems \cite{gompper-review,Bechingerreview}. 
For example, SPPs in hard confinements have been shown to accumulate at surfaces or walls \cite{elgeti2013wall} and 
collect in regions with higher concave curvature \cite{Fily_2014}, elucidating how the presence of  activity leads to different spatial distributions compared to thermal, Brownian systems \cite{bonn-softmat-review}. However, biological cells represent 
soft confinements, in which localized forces can deform the boundaries. This may create a complex feedback loop between boundary- and curvature-induced SPP accumulation, active force generation, and subsequent dynamic 
changes of vesicle shape. The possible occurrence of such a feedback loop raises questions about stationary and dynamic shapes, their transformation and dependence on the number of SPPs and their activity. 
Simulation studies of SPPs enclosed in semi-flexible polymer rings in two spatial dimensions (2D)
predict for high volume fractions of SPPs strong, predominantly prolate vesicle deformations and an 
inhomogeneous distribution of SPPs with accumulation at the poles \cite{Paoluzzi_2016,wang2019shape}.
Similarly, simulation studies of SPPs confined in 3D vesicles at fixed reduced volume $v$, again for 
relatively high volume fractions of SPPs, 
predict swim pressure-induced shape changes between prolate, oblate, and stomatocyte shapes around
$v=0.6$, where the free energy of these shapes in equilibrium is nearly identical, so that the system 
is highly susceptible to perturbations \cite{Li2019shape}.  At high swim pressures, active pearling of the vesicles is predicted. 

We present here the first experimental realization and characterization, combined with an extensive systematic simulation 
study, of such an active vesicle system with pronounced activity-induced fluctuations and shape changes. We
employ self-propelled Janus particles driven by local chemical gradients 
and confined within lipid vesicles to investigate activity-induced vesicle shapes and fluctuations. A simple fluid membrane made with 1,2-dioleoyl-sn-glycero-3-phosphocholine  (DOPC) is used here, but the methodology can be extended to more complex lipid compositions. The main results of our combined experimental and numerical study are that particle self-propulsion 
induces a plethora of new vesicle shapes and fluctuations, ranging from quasi-spherical shapes with active 
fluctuations to highly branched, dendritic structures and bola shapes. These ``active vesicles" 
(i) constitute minimalistic model systems for understanding how internal forces deform fluid lipid membranes and take them far from equilibrium, and (ii) may contribute to elucidating the underlying mechanisms 
of shape generation in complex-shaped cells, such as neurons and glial cells 
\cite{goldman1990cyclic, mcallister2000cellular, scott2001dendrites,chung2005human}.


\section*{Results and Discussion}

\paragraph*{Activity-induced vesicle shapes.}
Self-propelled Janus colloids of diameter $\sigma$ =  1.0 $\mu$m, which have nonspecific  interactions with the membrane (both in absence and presence of the propulsion, 
see SI), are encapsulated in giant unilamellar DOPC vesicles (GUVs) of radius $R$ (see Fig.~S1). Internally driven self-propelled particles are force and torque free, and a balance between the propulsion and the drag force results in a propulsion velocity $v_p$, controlled by varying the concentration of hydrogen peroxide (${\rm H}_2 {\rm O}_2$). When an SPP reaches the vesicle surface, it pushes against the membrane (see Movie S1) and the membrane's response to deformation is now what opposes the propulsion force. This force is exerted on the membrane as long as the Janus particle does not reorient, which is governed by the rotational diffusion time $\tau_{\rm r}=1/D_{\rm r}$, 
with rotational diffusion coefficient $D_{\rm r}$. The resulting membrane deformation, the elastic restoring force, and
the emerging complex and dynamic vesicle shapes are controlled by three parameters: the 
propulsion velocity $v_{\rm p}$, the number $N_{\rm p,c}$ of SPPs, and the membrane tension $\lambda$. 
Three dimensionless numbers describe the relative magnitude of these parameters: the P\'eclet number ${\rm Pe}=v_{\rm p}\sigma/D_{\rm t}$, with a translational diffusion coefficient $D_{\rm t}$, which indirectly measures the strength of the propulsion force, the SPP volume fraction $\phi$, which identifies the number of point forces, and a 
reduced tension $\gamma = \lambda R^2/k_BT$, which partially controls the membrane response.

\begin{figure}
\centering
\includegraphics[width=1.0\textwidth]{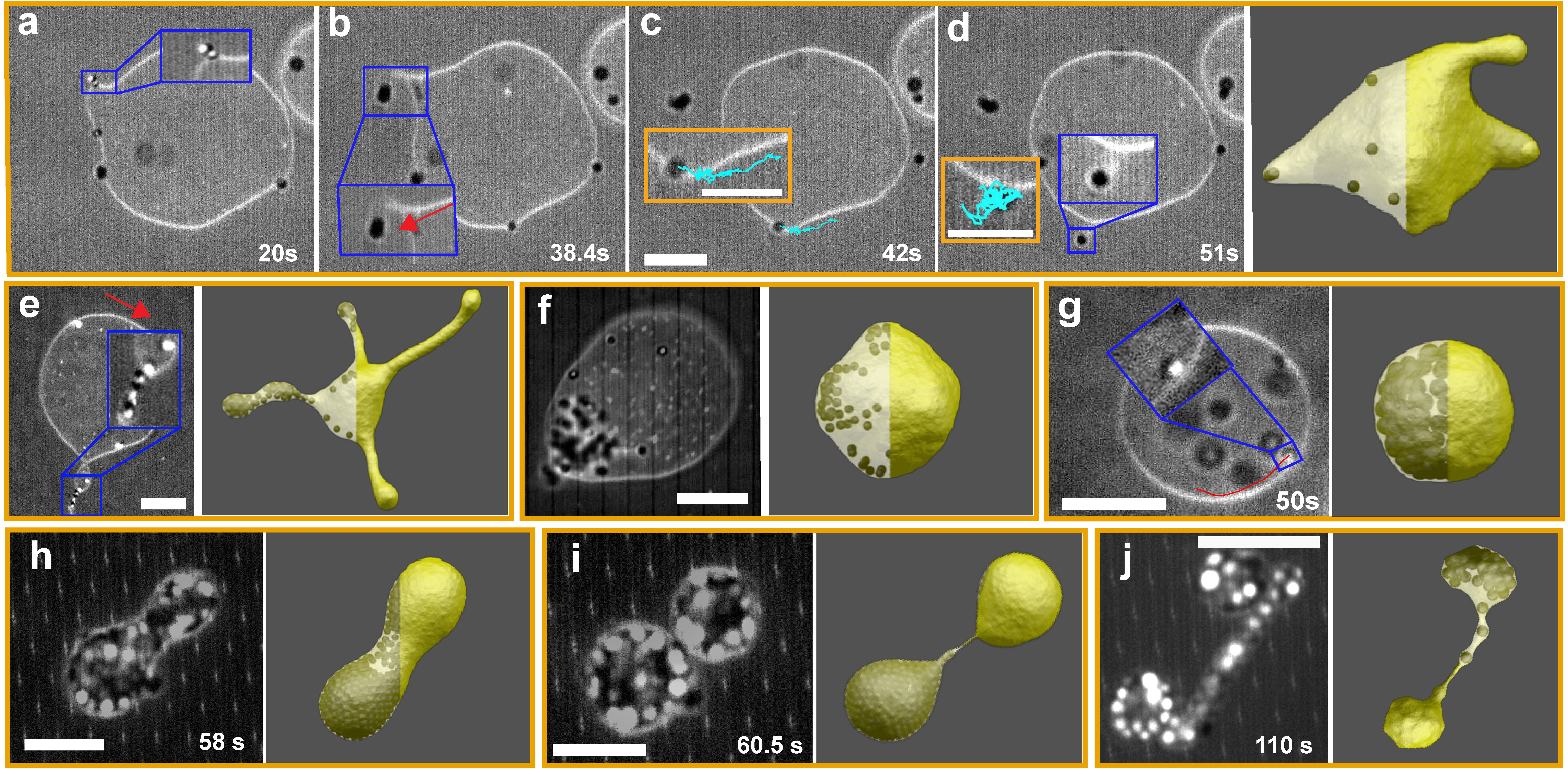}
\caption{{\bf Dynamic response of the lipid membrane at a low (flaccid) (a-f, h-j) and high (g) tension to the encapsulated SPPs at different densities $\phi$}. 
({\bf a-d}) Sequence of time-lapse combined 2D confocal and bright-field microscopy images showing various membrane deformations from partial and full wrapping 
to tethering and to channeling process. ({\bf a}) Tether and channel formation. The inset of ({\bf b}) shows that a second particle gets trapped in the channel. 
({\bf c-d}) Other SPP protrusions with partially and fully wrapped states, and subsequent budding formation (the inset in ({\bf d})). 
({\bf e}) Tethers transform into channels by multiple active particles. The inset illustrates a cooperative packing of SPPs. 
({\bf f}) Membrane curvature-mediated accumulation of SPPs. ({\bf g}) Dynamics of active particles in a high-tension vesicle, showing no tethering and 
the circular trajectory of an SPP near the membrane. ({\bf h-j}) Time-lapse images illustrating the shape change from a sphere to two twin vesicles connected 
by a tether. ({\bf b,e-j}) Simulation snapshots representing qualitatively the experimental observations.  
All scale bars are 10 $\mu$m. }
\label{fig1}
\end{figure}

A series of experiments was performed to interrogate membrane dynamics and shape transitions, for two values of the membrane tension $\lambda$ and various particle concentrations $\phi$ to increase the overall forces. The 
bending rigidity was estimated to be  $\kappa_{\rm c} = 15 \pm 5~k_{\rm B}T$ from the equilibrium fluctuation spectra 
(see SI).  The density of the internal fluid is higher than that of the external fluid causing the vesicles 
to sediment to the bottom of the observation chamber, where they become flattened and can easily be imaged 
by high-speed confocal and bright-field microscopy. Figure 1a-d shows time-lapse images of vesicles for a low particle concentration (${\phi} = 4\times 10^{-4}$) at low tension ($\lambda \sim 10 ~{\rm nN/m}$). Even though the propulsion forces $f_{\rm p}\approx0.1~{\rm pN}$ (for $v_{\rm p} =15.0 \pm 2.0 \, {\rm \mu m/s}$, ${\rm Pe} = 33-39$) of single SPPs are very small, they are strong enough to locally deform the membrane. Tether formation is typically initiated cooperatively by a 
few SPPs (see inset of Fig.~\ref{fig1}a).  As a few SPPs deform the membrane (Fig.~\ref{fig1}b),  additional SPPs  can 
become trapped in this region of higher curvature, thereby increasing the local force on the membrane and forming a tether (see also Movie S1).
A high degree of wrapping of SPPs by the membrane slows down particle movement significantly 
(see overlaid particle trajectories in insets of Figs.~\ref{fig1}c-d), most likely because 
partial or full wrapping of SPPs reduces their accessibility to the fuel and therefore their activity, as well as the viscoelastic
resistance due to membrane deformation.
In contrast, for high membrane tension ($\lambda  > 10 ~{\rm \mu N/m}$), the same propulsion force is 
too weak to overcome the deformation-energy cost for tether formation, as shown 
in Fig.~\ref{fig1}g. In this case, SPPs follow the membrane curvature and perform a circling motion (see Movie S1) similar to rodents in a hamster wheel. 

Higher particle concentrations ($\phi = 3 \times 10^{-3}$) in low-tension vesicles lead to the formation of long tubular protrusions where SPPs become tightly packed, as illustrated
in Fig.~\ref{fig1}e. Here, only particles that are close to the neck of a tube can still change their propulsion direction via rotational diffusion 
and move in or out of the protrusion (inset of Fig.~\ref{fig1}e and Fig.~S2). Changes in local membrane curvature induce pronounced clustering of SPPs and can even lead to
a cooperative wrapping process. For example, a cluster of several particles generates an increased local membrane stress, so that very long tethers ($\sim$200 {\rm$\mu$m}) 
can form (see Fig.~S3), which remain stable until particles reorient and tether retraction occurs.  

If the SPP concentration is sufficiently high, tether and protrusion formation are no longer observed. Figure \ref{fig1}f demonstrates near-spherical shapes for high SPP concentrations ($\phi= 10^{-2}$), where large SPP clusters, an accumulation of SPPs in regions of high membrane curvature, and rather global membrane deformations are observed. There are two possible explanations 
for the suppression of tethering and the new regime where strong global shape changes dominate: (i) the net propulsion force of the individual particles decreases because of crowding and their tendency to cluster due to diffusiophoretic attraction \cite{Bechingerreview,gompper-review}, and (ii) particle-induced membrane tension prevents large local deformations of the vesicle 
(see state diagram and shape fluctuations analysis section). At very high SPP concentrations, e.g. $\phi = 1.9 \times 10^{-1}$, 
the vesicle can transform into bola-like and prolate shapes (Figs.~\ref{fig1}h-i and  Movie S2),
with satellites connected by thin particle-laden tethers (Fig.~\ref{fig1}j).  

We employ computer simulations in addition to experiments to explore the parameter space in more 
detail, in particular the role of P\'eclet number, and to elucidate the underlying mechanisms. The {\em in silico} 
model consists of active Brownian particles characterized by a propulsion velocity $v_p$ and an orientational
diffusion time $\tau_r$, enclosed by a dynamically triangulated surface \cite{Kroll1992a}, 
representing the membrane. Bending elasticity, the membrane fluidity of the DOPC membranes and area constraint are taken into account, 
sometimes in combination with a volume constraint (corresponding to fixed reduced volume)  (see SI for details).


\paragraph{State diagram of active-vesicle shapes.}

In our computer simulations, the focus is on flaccid vesicles (no volume constraint) where the most dramatic changes 
are observed experimentally. In this regime, the membrane properties are predominantly determined by bending elasticity. 
Selected experimental and simulation snapshots are shown in Fig.~\ref{fig1} to demonstrate that the 
observed structures, i.e., tethers (b), dendrites (e), and bolas and dumbbells (h-j), agree very well 
when shapes are compared at similar P\'eclet numbers, although particle densities are sometimes slightly different. Movies S3-S5 
showcase dynamic formation of a few vesicle shapes for different Pe and $\phi$. 

\begin{figure}
\centering
\includegraphics[width=0.75\textwidth]{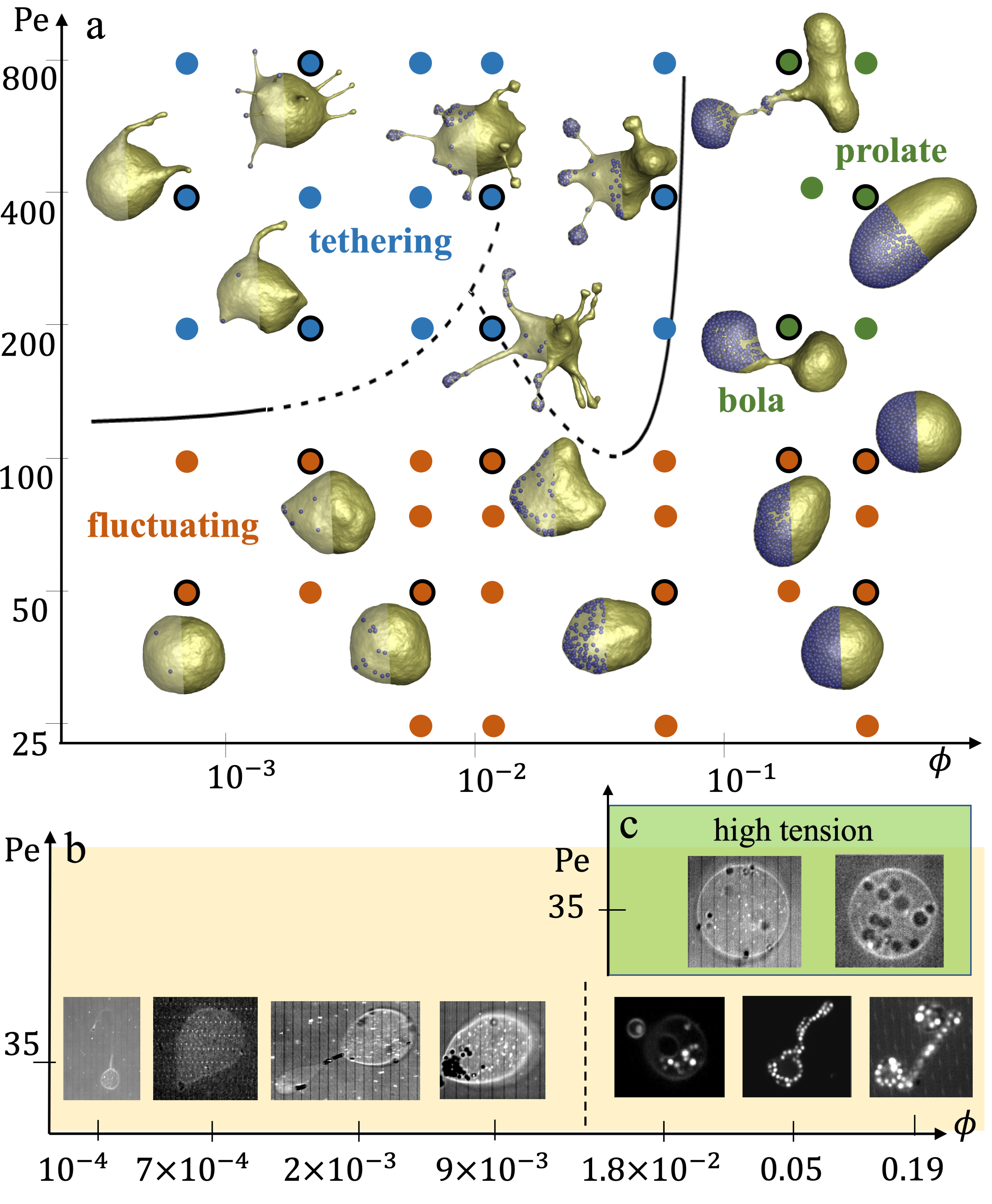}
    \caption{ {\bf State diagram of various membrane structures}. ({\bf a}) Simulated diagram for different $\rm Pe$ and $\phi$ with the three regimes:
    	tethering (blue symbols), fluctuating (red symbols), and bola/prolate (green symbols) vesicle shapes. The points corresponding to the snapshots are marked by open black circles.
    	All simulations are without volume constraint, mimicking a flaccid vesicle. 
	 ({\bf b}) The experimental diagram summarizes the observed vesicle shapes 
    	induced by the activity of SPPs for different $\phi$ and a fixed $\rm Pe$. ({\bf c}) A high-tension vesicle with no significant deformations.}
    \label{fig2}
\end{figure}

The full state diagram of flaccid active vesicles as function of particle P\'eclet number $\rm Pe$ 
and volume fraction $\phi$ is shown in Fig.~\ref{fig2}a. Three distinct regimes are identified: (i) The 
``dendritic tether'' regime with long and thin tethers originating from a mother vesicle occurs for only large $\rm Pe$ and small to moderate $\phi$. With increasing $\phi$, the number of tethers first increases, 
as each SPP pulls its own tether, and then decreases as several SPPs tend to cluster in one tether. 
An interesting feature is the filling of the tethers by additional SPPs and the formation of large 
SPP clusters in small satellite vesicles at the end of the tethers. (ii) The ``fluctuating vesicle" regime
is characterized by the absence of tethers and quasi-spherical vesicle shapes, which is found for sufficiently small values of $\rm Pe$.  (iii) The ``bola/prolate" regime 
occurs for large $\rm Pe$ and $\phi$ values and displays strong global shape changes.


For small P\'eclet number ${\rm Pe}$, the particles reorient before they can deform 
the membrane, whereas for large ${\rm Pe}$ tethers form.
Tether formation occurs when the propulsion force ($f_{\rm p}$) 
exceeds the force required to deform the membrane, similarly to classical tether-pulling experiments 
\cite{Heinrich_1996}.
A theoretical estimate for the phase boundaries between the three regimes can be obtained as follows. 
The propulsion force $f_{\rm p}$ written as a function of $\rm Pe$ is given by
$ f_{\rm p} = v_{\rm p} \gamma_{\rm p} = v_{\rm p}k_{\rm B}T/D_{\rm t} = {\rm Pe}\,k_{\rm B}T/\sigma$, where $\gamma_{\rm p}$ is the translational friction coefficient and $k_{\rm B}T$ is the thermal energy.
The membrane may form small satellite vesicles containing a cluster of $N_{\rm p,c}$ SPPs pushing cooperatively,
which generate a net maximum force of $f_{\rm p}N_{\rm p,c}$. 
For the case of flaccid vesicles,
the resistance to bending can be approximated as $f_{\rm bend} = 2 \pi \kappa_{\rm c} / w_{\rm teth}$ for a tether
of thickness $w_{\rm teth}=\sigma$ and bending rigidity $\kappa_{\rm c}$ \cite{peterson2008random,phillips2012physical}.
Therefore, the critical $\rm Pe_c$ is 
$   {\rm Pe}_{\rm c} N_{\rm p,c}  = 2\pi \kappa_{\rm c} / {k_{\rm B}T}  $, 
beyond which the vesicle forms tethers (see SI for details). For a typical value of the bending rigidity, 
i.e.~$\kappa_{\rm c} = 20k_{\rm B}T$, this estimate yields ${\rm Pe}_{\rm c} \simeq 125$ for a single SPP, which agrees 
well with the simulated phase boundary in Fig.~\ref{fig2}a. 

The generalization of this estimate to the case with intrinsic or activity-induced membrane tension (see SI) yields
\begin{equation}
Pe_c = \frac{2\pi \kappa}{N_{p,c}} \left( 1 - \frac{N_p}{8}\frac{\sigma}{R} \right)^{-1}.
\end{equation}
With increasing total number $N_{\rm p}$ of particles, both the total internal swim pressure and the induced membrane tension increase,
which suppresses tether formation. 
The boundary between the tethering regime and the bola/prolate regime is defined by the suppression of tether formation by 
activity-induced membrane tension, see Fig.~\ref{fig2}a. 
This boundary is consistent with the simulations: at high $\phi$, the SPPs form a thick particle layer which exerts a nearly homogeneous swim pressure 
\cite{Takatori_SPG_2014} on the membrane, so that membrane 
deformations are strongly suppressed.  Interestingly, in the ``bola/prolate" regime, 
this also leads to the formation of two or more satellite vesicles.


Figure \ref{fig2}b demonstrates that the experimental results are in qualitative agreement with simulations in the 
limit of a tensionless vesicle. 
In addition, experiments reveal changes in topology, from extended structures to donut-shaped, stomatocyte vesicles 
(see Figs.~\ref{fig2}b and S5). This is not observed in the simulations, because topology changes are not allowed (see SI). 
At high particle loadings, vesicle shapes are not governed by intrinsic passive tension, 
which is small for flaccid vesicles, but by active tension generated by the SPPs stretching the membrane. 
The active tension in flaccid vesicles acts in a similar way as a large enough passive tension (Fig.~\ref{fig2}c), where 
the resistance to membrane deformation increases as $f_{\rm tense} = \pi w_{\rm teth} \lambda$. 
Figure~\ref{fig2}c shows that for high passive membrane tension ($\sim$ 20 $\mu$N/m), 
no visible effects of activity on the macroscopic vesicle shape are observed under conditions where flaccid vesicles 
would show dendritic structures. 
It is interesting to note that in
our out-of-equilibrium system, shape and tension can vary dynamically, as demonstrated by 
Movie S6, where a cluster of fast moving ($v_p \sim 21 \mu$m/s) SPPs leads to the formation of a daughter vesicle,
resulting in membrane stiffening by high induced tension in the mother vesicle (see SI).

Despite substantial simplifications in the simulation model, where details of 
the diffusiophoretic particle propulsion and hydrodynamic 
interactions are omitted, the general characteristics of the system are well captured. 
However, for equivalent vesicle conformations, the P\'eclet numbers in experiment are typically smaller 
by a factor of 4-5 than in simulations. This could be due to a local spontaneous membrane 
curvature induced by a local change in lipid concentration, to differences in membrane viscoelasticity, or to
changes in reorientation time because of particle wrapping by the membrane in the experimental system (see SI).


\paragraph*{Shape fluctuations.}

\begin{figure}

\centering
\includegraphics[width=\textwidth]{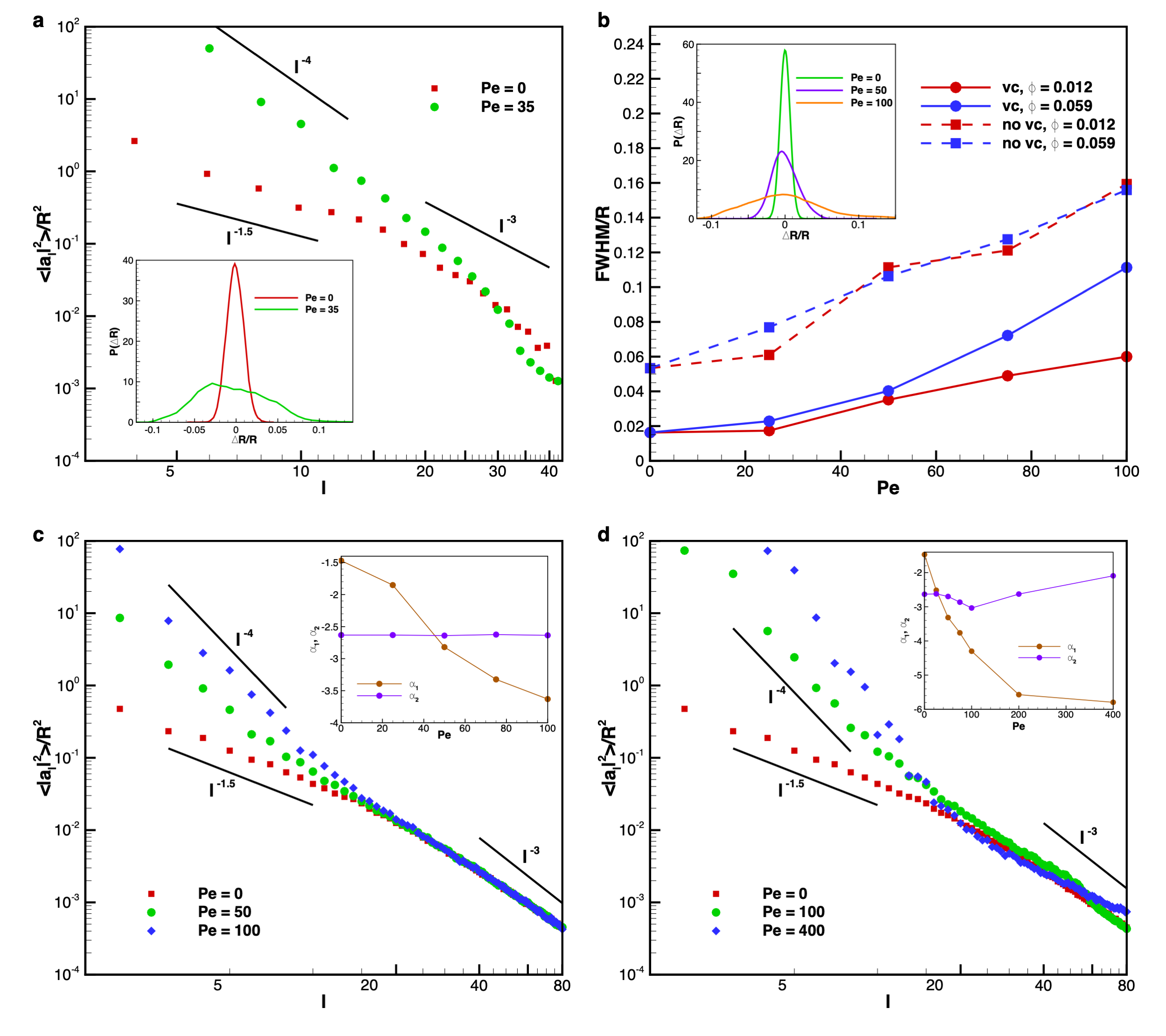}
\caption{{\bf Shape fluctuation spectra for both passive and active vesicles}. ({\bf a}) Experimental shape fluctuation spectra as a function of the mode number $\ell$ for GUVs that 
	enclose passive and active colloids with volume fraction $\phi \approx 0.009$. Inset: Experimental probability density functions (PDF) of the radial position of the fluctuating 
	contour from its average position for low-tension passive and active vesicles, respectively. ({\bf b}) Normalized FWHM (full width at half maximum) for different Pe and $\phi$ 
	with and without volume constraint (vc) from 3D simulations. Inset: PDFs of the radial position of the fluctuating contour from its average position for the case with vc ($0.96< v < 0.99$) and 
	$\phi = 0.059$. 
({\bf c-d}) Simulation spectra for vesicles with vc ($0.96< v < 0.99$) for volume fractions ({\bf c}) $\phi = 0.059$  and ({\bf d}) $\phi = 0.356$ and various $\rm Pe$ numbers.
The insets show the exponents obtained from power-law fits to $\ell^{\alpha}$, measured in the ranges $3 \leq \ell \leq 12$  for $\alpha_1$ and $35 \leq \ell \leq 80$ for $\alpha_2$.}
\label{fig3}
\end{figure}

To obtain a better understanding of how the active particles influence membrane dynamics, the 
spectra of undulation modes are determined from 2D equatorial contours in the ``fluctuating vesicle" regime, both 
from experiments and simulations. 
Figure \ref{fig3}a demonstrates that the experimental fluctuation amplitudes are strongly enhanced due to activity compared
to the passive case. This is reflected both in larger mode amplitudes as well as an increased width of the 
probability density function (PDF) $\mathcal{P}(\Delta r)$ of the 
distance between the instantaneous membrane contour and its average position
(see inset of Fig.~\ref{fig3}a).

Corresponding simulation results for various activities and numbers of SPPs are shown in Fig.~\ref{fig3}b-d. 
For a vesicle under quiescent conditions, the Helfrich curvature elasticity Hamiltonian 
predicts a Gaussian PDF, with an FWHM (full width at half maximum) determined by bending rigidity and membrane tension. 
The simulation results in Fig.~\ref{fig3}b demonstrate that the FWHM of the PDF increases roughly linearly with $Pe$. 
An interesting difference is
observed in the fluctuation magnitudes of vesicles with and without volume constraint (Fig.~\ref{fig3}b). Here, the smaller 
fluctuations with volume constraint can be traced back to the intrinsic membrane tension 
of a vesicle with reduced volume $v$ close to unity.

Full spectra of undulation amplitudes $\langle a_\ell^2 \rangle$ versus mode numbers $\ell$ are
displayed in Figs.~\ref{fig3} a, c, and d. Experimental and simulation results
show very good agreement of all salient features. The spectra are characterized by 
two power-law regimes $\ell^{\alpha}$. For passive vesicles with volume constraint, we find 
the exponent $\alpha\simeq -1$ for small $\ell$ (tension-dominated regime), and $\alpha=-3$ 
for larger $\ell$ (bending-dominated regime), in agreement with earlier observations \cite{Bassereau2004}. 
For active vesicles, the activity leads to a very pronounced enhancement of the small-$\ell$ modes, with the
exponent $\alpha\simeq-4$, whereas the bending-dominated regime remains essentially unaffected.
Another important feature is the dependence on the particle volume fraction $\phi$. With increasing
$\phi$, the fluctuation amplitudes increase (see Fig.~\ref{fig3}b), which is also reflected in the
more negative exponent $\alpha$ in the fluctuation spectrum (see Fig.~\ref{fig3}d). This happens,
because the density and thickness of the particle layer increase, generating a larger active force
on the membrane. However, this increase is moderate, as it is opposed by a larger induced membrane
tension.

The effect of activity on membrane fluctuations has previously been studied experimentally for vesicles with 
light-activated membrane pumps \cite{mannevilleprl} and for red blood cells  
with ATP-dependent cytoskeletal activity \cite{park2010}.
The results for these very different active-membrane systems, including ours, all agree on an enhancement of 
membrane undulations due to activity. Analytical calculations for quasi-spherical vesicles \cite{loubet2012} and 
polymer rings \cite{mousavi2019} 
with spatially uncorrelated colored noise (mimicking the activity) predict a pronounced power-law increase 
of the fluctuation spectrum for small modes numbers $\ell$.  
We now see this enhancement for the first time in our entirely synthetic active particle-membrane 
system.

\section*{Conclusions}
We have presented an artificial biomimicking system, consisting of a giant unilamellar vesicle which encapsulates 
a number of self-propelled particles. The active particles generate local active forces from the inside, which
implies strong out-of-equilibrium shape deformations.
Striking tethered structures are observed, especially at low particle concentrations. Regions of high local 
curvature lead to SPP accumulation or clustering, which further enhances tether formation. Vesicle loading 
with a high particle concentration changes the deformations from local to collective effects, and to global shape 
distortions. The analysis of membrane fluctuations demonstrates a strong departure from thermal fluctuations governed by the Helfrich curvature-elasticity 
model when active particles are present, emphasizing the importance of active non-equilibrium processes. Our simulations 
agree well with the experimental observations, and enable the construction of a vesicle state diagram for a wide range 
of SPP activities and concentrations.
Strikingly - less is more - the most dramatic shape changes are observed at low particle concentrations, as is evidenced 
by a state diagram. Our results provide a framework for understanding the conditions under which cells may control 
their shape. Moreover, further analysis of the strongly distorted vesicles shapes may provide a framework to study 
non-equilibrium thermodynamic and rheological responses of bilayer membranes when the membrane composition becomes 
more complex.



\section*{Acknowledgments}
 H.R.V. acknowledges financial support
through a Marie Sk{\l}odowska-Curie Intra European Individual Fellowship (G.A. No. 708349- SPCOLPS) within Horizon 2020. J.V.
acknowledges financial support by the Swiss National Science Foundation. C.A.V. acknowledges support by the International Helmholtz
Research School of Biophysics and Soft Matter (IHRS BioSoft). We would like to thank Prof. Peter Walde (ETH, Z\"urich, Switzerland) for useful discussions and Nick Jaensson (ETH, Z\"urich, Switzerland) 
is acknowledged for help with the membrane curve fitting of the experiment. We also gratefully acknowledge the computing time granted through
JARA-HPC on the supercomputer JURECA  at Forschungszentrum J\"ulich. 

\section*{Author contributions}
H.R.V. and J.V. conceived and designed the project. H.R.V., B.v.L., and  A.D. performed the experimental work. H.R.V. performed the experimental analysis. 
T.A., D.A.F., and G.G. designed the simulations. M.H. and D.A.F. performed 3D simulations. C.A.V performed 2D simulations. M.H., C.A.V., and D.A.F. analyzed the simulation data. 
M.H., C.A.V, and T.A. performed the analytical calculations. H.R.V., M.H., C.A.V., T.A., D.A.F., G.G., and J.V. participated in the discussions and wrote the manuscript.

\section*{Competing interests}
The authors declare no competing financial interests.

\section*{Data and materials availability}
The data that support the findings of this study are available from the corresponding authors upon reasonable request.

\newpage

\begin{thebibliography}{10}

\bibitem{Mattila_FMA_2008}
P.~K. Mattila, P.~Lappalainen, {\it Nat. Rev. Mol. Cell Biol.\/} {\bf 9}, 446
  (2008).

\bibitem{Fletcher_CMC_2010}
D.~A. Fletcher, R.~D. Mullins, {\it Nature\/} {\bf 463}, 485 (2010).

\bibitem{needleman2017active}
D.~Needleman, Z.~Dogic, {\it Nat. Rev. Mater.\/} {\bf 2}, 17048 (2017).

\bibitem{turlier2016}
H.~Turlier, {\it et~al.\/}, {\it {Nat. Phys.}\/} {\bf 12}, 513 (2016).

\bibitem{Bausch-science}
F.~C. Keber, {\it et~al.\/}, {\it Science\/} {\bf 345}, 1135 (2014).

\bibitem{Dogtram-cell}
L.~Laan, {\it et~al.\/}, {\it Cell\/} {\bf 148}, 502  (2012).

\bibitem{Mulla_2018}
Y.~Mulla, A.~Aufderhorst-Roberts, G.~H. Koenderink, {\it Phys. Biol.\/} {\bf
  15}, 041001 (2018).

\bibitem{petra-review2019}
K.~A. Ganzinger, P.~Schwille, {\it J. Cell Sci.\/} {\bf 132}, jcs227488 (2019).

\bibitem{mellouli2013self}
S.~Mellouli, {\it et~al.\/}, {\it Soft Matter\/} {\bf 9}, 10493 (2013).

\bibitem{Andriezen_NGE_1893}
W.~L. Andriezen, {\it Br. Med. J.\/} {\bf 2}, 227 (1893).

\bibitem{Vasile_HAC_2017}
F.~Vasile, E.~Dossi, N.~Rouach, {\it Brain Struct. Funct.\/} {\bf 222}, 2017
  (2017).

\bibitem{theriot1992rate}
J.~A. Theriot, T.~J. Mitchison, L.~G. Tilney, D.~A. Portnoy, {\it Nature\/}
  {\bf 357}, 257 (1992).

\bibitem{gompper-review}
J.~Elgeti, R.~G. Winkler, G.~Gompper, {\it Rep. Prog. Phys.\/} {\bf 78}, 056601
  (2015).

\bibitem{Bechingerreview}
C.~Bechinger, {\it et~al.\/}, {\it Rev. Mod. Phys.\/} {\bf 88}, 045006 (2016).

\bibitem{elgeti2013wall}
J.~Elgeti, G.~Gompper, {\it EPL (Europhys. Lett.)\/} {\bf 101}, 48003 (2013).

\bibitem{Fily_2014}
Y.~Fily, A.~Baskaran, M.~F. Hagan, {\it Soft Matter\/} {\bf 10}, 5609 (2014).

\bibitem{bonn-softmat-review}
R.~W. Jaggers, S.~A. Bon, {\it Soft Matter\/} {\bf 14}, 6949 (2018).

\bibitem{Paoluzzi_2016}
M.~Paoluzzi, R.~{Di Leonardo}, M.~C. Marchetti, L.~Angelani, {\it Sci. Rep.\/}
  {\bf 6}, 34146 (2016).

\bibitem{wang2019shape}
C.~Wang, Y.-k. Guo, W.-d. Tian, K.~Chen, {\it J. Chem. Phys.\/} {\bf 150},
  044907 (2019).

\bibitem{Li2019shape}
Y.~Li, P.~ten Wolde, {\it Phys. Rev. Lett.\/} {\bf 123}, 148003 (2019).

\bibitem{goldman1990cyclic}
J.~E. Goldman, B.~Abramson, {\it Brain Res.\/} {\bf 528}, 189 (1990).

\bibitem{mcallister2000cellular}
A.~K. McAllister, {\it Cereb. Cortex\/} {\bf 10}, 963 (2000).

\bibitem{scott2001dendrites}
E.~K. Scott, L.~Luo, {\it Nat. Neurosci.\/} {\bf 4}, 359 (2001).

\bibitem{chung2005human}
B.~G. Chung, {\it et~al.\/}, {\it Lab Chip\/} {\bf 5}, 401 (2005).

\bibitem{Kroll1992a}
D.~Kroll, G.~Gompper, {\it Science\/} {\bf 255}, 968 (1992).

\bibitem{Heinrich_1996}
V.~Heinrich, R.~E. Waugh, {\it Ann. Biomed. Eng.\/} {\bf 24}, 595 (1996).

\bibitem{peterson2008random}
E.~L. Peterson, {\it A random walk in physical biology\/} (California Institute
  of Technology, 2008).

\bibitem{phillips2012physical}
R.~Phillips, J.~Kondev, J.~Theriot, H.~Garcia, {\it Physical Biology of the
  Cell\/} (Garland Science, 2012).

\bibitem{Takatori_SPG_2014}
S.~C. Takatori, W.~Yan, J.~F. Brady, {\it Phys. Rev. Lett.\/} {\bf 113}, 028103
  (2014).

\bibitem{Bassereau2004}
J.~P{\'e}cr{\'e}aux, H.-G. D{\"o}bereiner, J.~Prost, J.-F. Joanny,
  P.~Bassereau, {\it Eur. Phys. J. E\/} {\bf 13}, 277 (2004).

\bibitem{mannevilleprl}
J.-B. Manneville, P.~Bassereau, D.~L\'evy, J.~Prost, {\it Phys. Rev. Lett.\/}
  {\bf 82}, 4356 (1999).

\bibitem{park2010}
Y.~Park, {\it et~al.\/}, {\it Proc. Natl. Acad. Sci. U.S.A.\/} {\bf 107}, 1289
  (2010).

\bibitem{loubet2012}
B.~Loubet, U.~Seifert, M.~A. Lomholt, {\it Phys. Rev. E\/} {\bf 85}, 031913
  (2012).

\bibitem{mousavi2019}
S.~M. Mousavi, G.~Gompper, R.~G. Winkler, {\it J. Chem. Phys.\/} {\bf 150},
  064913 (2019).

\end{thebibliography}
\end{document}